\documentclass[]{JHEP3}
\usepackage{epsfig}

\newcommand{\vub}{V_\mathrm{ub}}
\newcommand{\GF}{G_\mathrm{F}}
\newcommand{\order}[1]{{\mathcal O}(#1)}
\newcommand{\pme}[2]{{}^{+#1}_{-#2}}
\newcommand{\psb}{\bar{\psi}}

\newcommand{\cl}[1]{\mathcal{#1}}
\newcommand{\vk}{v\cdot k}
\newcommand{\mev}{\,\mathrm{MeV}}
\newcommand{\gev}{\,\mathrm{GeV}}
\newcommand{\fm}{\,\mathrm{fm}}

\newcommand{\ba}{\begin{eqnarray}}
\newcommand{\ea}{\end{eqnarray}}
\newcommand{\be}{\begin{equation}}
\newcommand{\ee}{\end{equation}}

\newcommand{\GammaPI}{%
4.9^{+12}_{-10}{}^{+\ 0}_{-14} \times  10^{12}\,\mathrm{s}^{-1}}

\title{B $\to\rho l\nu$ form factors in lattice QCD}
\author{UKQCD Collaboration}
\author{K.C.~Bowler, J.F.~Gill, C.M.~Maynard, \\ School of Physics, University of Edinburgh,
 Edinburgh, EH9 3JZ, UK}
\author{J.M.~Flynn\\School of Physics \& Astronomy, University of Southampton,
        Southampton, SO17 1BJ, UK}
\abstract{We present results from quenched lattice QCD for the
form factors for the decay $B\to\rho l\nu$. The calculations are
performed using a nonperturbatively improved action and operators
at two values of the lattice spacing. The bottom quark mass is
reached by extrapolation from simulations performed with heavy
quark masses around the charm mass. Our primary result is for the
partially integrated decay rate $\Gamma_\mathrm{PI}$ over the
range $12.7\gev^2 < q^2 < 18.2\gev^2$:
 \[
 \Gamma_\mathrm{PI} = \GammaPI\,|\vub|^2.
 \]}
 \preprint{Edinburgh-2004/04, SHEP--0404}

\begin{document}

\section{Introduction}

One of the primary goals of modern particle physics experiments is
to determine the elements of the Cabibbo-Kobayashi-Maskawa (CKM)
matrix. This matrix is unitary in the Standard Model (SM) and
so-called unitarity triangles can be constructed using the
orthogonality of pairs of rows or columns. For the most common
triangle, drawn in the complex $\rho$--$\eta$ plane of the
Wolfenstein parameterisation of the CKM matrix~\cite{wolfenstein},
BaBar and Belle have produced the most accurate measurements of
one of the angles (via $\sin 2\beta$) to
date~\cite{Belle_s2b_2003,BaBar_s2b_2002}. To fully constrain the
triangle one needs to know the length of the opposing side,
governed by $\vub$. Any process in which a $b$ quark transforms to
a $u$ quark can be used to measure $\vub$. Two possibilities are
the exclusive decays $B\to\pi l\nu$ and $B\to\rho l \nu$. The form
factors which parameterise the hadronic amplitudes for these
decays can be determined from first principles in lattice QCD.
Combining these with measurements of the decay rate would allow
the determination of $\vub$. In particular, combining a measured
differential decay rate with the lattice values for the form
factors at high $q^2$ would allow $\vub$ to be extracted
independently of any model for the $q^2$ behaviour of the form
factors.

The transition amplitude for semileptonic decays factorises into
leptonic and hadronic parts when $q^2 \ll m_W^2$, where $q^2$ is
the momentum transfer squared and $m_W$ is the mass of the $W$
boson. The hadronic matrix element contains the non-perturbative
strong interaction effects and is the largest source of
uncertainty in theoretical determinations of the decay rates.

The hadronic matrix elements for $B\to\rho l\nu$ are parameterised
by four form factors
 \ba
 \label{eqn:FF}
 \langle \rho(\vec{k}),\eta | V_{\mu} | B(\vec{p}) \rangle
  & = &
   \frac{2 V(q^2)}{m_B+m_\rho} \varepsilon_{\mu\rho\sigma\delta}
   p^{\rho}k^{\sigma} \eta^{\star \delta} \\\nonumber
 \langle \rho(\vec{k}),\eta | A_{\mu} | B(\vec{p}) \rangle
  & = &
    i(m_B+m_\rho)A_1(q^2) g_{\mu\sigma} \eta^{\star\sigma}_r
    -\frac{iA_2(q^2)}{m_B+m_\rho} (p+k)_{\mu}
     q_\sigma \eta^{\star\sigma}_r \nonumber \\
  && \mbox{}+
     \frac{2im_\rho A(q^2)}{q^2}(p-k)_{\mu}(p+k)_\sigma \eta^{\star\sigma}_r
 \ea
where $B$ is the initial state pseudoscalar meson with
three-momentum $\vec{p}$, $\rho$ is the final state vector meson
with three-momentum $\vec{k}$ and polarisation $\eta$, and
$q=p-k$. The form factor $A$ can be written as
 \be
 \label{eqn:A_def}
  A(q^2)=A_0(q^2)-A_3(q^2)
 \ee
where
 \be
  A_3(q^2)=\frac{m_B+m_\rho}{2m_\rho}A_1(q^2) - \frac{m_B-m_\rho}{2m_\rho}A_2(q^2)
 \ee
with $A_0(0)=A_3(0)$.

In the limit of zero lepton mass the differential decay rate is
given by~\cite{korner_schuler_dr}
 \be \label{eqn:ddr_V}
 \frac{d\Gamma}{dq^2} \ =\ \frac{\GF^2 | V_{\rm u b} |^2 }{192
 \pi^3 m_B^3} q^2 [\lambda(q^2)]^{\frac{1}{2}}\ \Big(\ |
 H^0(q^2)|^2 + | H^+(q^2)|^2 + | H^-(q^2)|^2 \ \Big)
 \ee
where
 \ba
 H^0(q^2)&=&\frac{1}{2m_\rho\sqrt{q^2}}\left\{
 \frac{4m_B^2|\vec{k}|^2}{m_B+m_\rho} A_2(q^2)
 - \left(m_B^2 -m_\rho^2
 -q^2\right)\left(m_B+m_\rho\right)A_1(q^2)\right\}, \\
 H^{\pm}(q^2)&=&\left(m_B+m_\rho\right)A_1(q^2) \pm
 \frac{2m_B | \vec{k} |}{m_B+m_\rho} V(q^2).
 \ea
$\GF$ is the Fermi constant and $\lambda$ is the kinematic factor,
 \be
  \lambda(q^2) =
  \left(m_B^2+m_\rho^2-q^2\right)^2 - 4m_B^2m_\rho^2
 \ee
$\vub$ can be extracted from the experimental decay rate once the
theoretical rate has been determined. The form factors $H_0$ and
$H_\pm$ correspond to the contributions of longitudinally and
transversely polarised light vector mesons
respectively~\cite{gilman_singleton}.

We present results for the form factors of the decays $B\to \rho
l\nu$ determined in the quenched approximation to lattice QCD, at
two values of the lattice spacing. We use the non-perturbatively
(NP) improved Sheikholeslami-Wohlert (SW)~\cite{sw_paper} action
and improved operators. Lattice artefacts then appear formally at
$\order{a^2}$ (where $a$ is the lattice spacing) rather than
$\order{a}$, though this does not guarantee that they are
numerically small.

Mass-dependent lattice artefacts depending on $(am_Q)^2$ in the NP
improved renormalisation scheme limit the heavy quark mass $m_Q$ at
which one can simulate, without resorting to an effective action such
as the Fermilab formalism~\cite{KLM_norm}. We have simulated heavy
quarks with several different masses around the charm quark mass, then
used continuum Heavy Quark Symmetry (HQS) to motivate the form of our
extrapolation to the $b$ quark mass. In this way $am_Q$ and thus
$(am_Q)^n$ are less than one. We have already presented a
determination of the form factors for $B\to \pi l\nu$ at our finest
value of the lattice spacing~\cite{B2pi_plb}.

The rest of the paper is organised as follows. Section two
describes the details of the calculation and how the form factors
were extracted from the data and then interpolated and
extrapolated to physical quark masses. Section three discusses the
results and contains an analysis of systematic errors.

The results reported here represent a new analysis of the data
used for an earlier preliminary analysis
in~\cite{ukqcd_gill_brho}.

\section{Details of the calculation}
\label{sec:details}

\subsection{Improved action and operators}
\label{sec:imp-action-ops}

In the Wilson formulation of lattice QCD, the fermionic part of the
action has lattice artefacts of ${\mathcal{O}}(a)$, while the gauge
action differs from the continuum Yang-Mills action by terms of
${\mathcal{O}}(a^{2})$.  To leading order in $a$ the Symanzik
improvement program involves adding the SW term to the fermionic
Wilson action,
 \begin{equation}
 S_{{\rm SW}}=S_{{\rm W}} - c_{{\rm SW}} \frac{i\kappa}{2}\sum_{x}
 {\bar \psi}(x) i \sigma_{\mu\nu} F_{\mu\nu}(x) \psi(x)
 \end{equation}
Full ${\mathcal{O}}(a)$ improvement of on-shell matrix elements also
requires that the currents are suitably improved. The improved vector
current is
 \begin{eqnarray}
 \label{eqn:imp_current}
 V_{\mu}^\mathrm{I}(x) & = & V_{\mu}(x)+ac_V{\tilde
                         \partial}_{\nu}T_{\mu\nu}(x) \nonumber \\
 A_{\mu}^\mathrm{I}(x) & = & A_{\mu}(x) +
                         ac_A{\tilde \partial}_{\mu}P(x)
 \end{eqnarray}
where
 \begin{eqnarray}
 V_{\mu}(x) & = & {\bar \psi}(x) \gamma_{\mu} \psi(x) \nonumber \\
 A_{\mu}(x) & = & {\bar \psi}(x) \gamma_{\mu}\gamma_{5} \psi(x)\nonumber \\
 P(x)       & = & {\bar \psi}(x) \gamma_{5} \psi(x) \nonumber \\
 T_{\mu\nu}(x) & = & {\bar \psi}(x)i\sigma_{\mu\nu} \psi(x) \nonumber
 \end{eqnarray}
and ${\tilde \partial_{\mu}}$ is the symmetric lattice derivative.
The current renormalisation is as follows $(J=A,V)$:
\begin{equation}
\label{eqn:renorm_current}
J^{\rm R} = Z_J(1+b_Jam_q)J^{\rm I}
\end{equation}
where $Z_J$ is calculated in a mass-independent renormalisation
scheme.

The bare quark mass, $am_q$, is
\begin{equation}
\label{eqn:amq}
am_q= \frac{1}{2} \left( \frac{1}{\kappa} - \frac{1}{\kappa_{\rm
crit}} \right)
\end{equation}
where $\kappa$ is the hopping parameter.  For non-degenerate currents,
an effective quark mass is used in the definition of the renormalised
current, corresponding to
\begin{equation}
\label{eqn:keff}
\frac{1}{\kappa_{\rm eff}}=
\frac{1}{2}\left(\frac{1}{\kappa_{1}}+\frac{1}{\kappa_{2}}\right)
\end{equation}
In this renormalisation scheme, the improved quark mass, used in the
chiral extrapolations, is defined as
\begin{equation}
\label{eqn:mimp}
{\widetilde m}_q = m_q (1+b_{\rm m} am_q)
\end{equation}

\subsection{Simulation details}

Gauge field configurations were generated using a combination of the
over-relaxed~\cite{creutz_or,brown_woch_or} and the
Cabibbo-Marinari~\cite{cabibbo_marinari} algorithms with periodic
boundary conditions at two values of the gauge coupling
$\beta=6/g_0^2$.  At each $\beta$, heavy quark propagators were
computed at four values of the hopping parameter, corresponding to
quarks with masses in the region of the charm quark mass.  For light
quarks, three values of $\kappa$ were used for the light quark which
occurs in the current, the active quark (A), and, owing to disk space
constraints, only two values of $\kappa$ for the passive quark
(P). All the light quarks had masses around that of the strange quark.
Table~\ref{tab:lattices} lists the input parameters.  \TABLE{
\caption{Input and derived parameters. The lattice spacing is set by
$r_0$.\smallskip} \label{tab:lattices} \begin{tabular}{lcc}
\hline\hline & $\beta=6.2$ & $\beta=6.0$ \\ \hline Volume &
$24^{3}\times 48 $& $16^{3}\times 48 $\\ $c_{\mathrm{SW}} $ & 1.614 &
1.769 \\ $N_{\mathrm{configs}}$ & 216 & 305 \\ ${a^{-1}}
\left({\mathrm{GeV}} \right) $ & $2.91^{+1}_{-1}$ & $2.12\pme{1}{1}$\\
Heavy $\kappa$ & $0.1200,0.1233,0.1266,0.1299$ &
$0.1123,0.1173,0.1223,0.1273$\\ Light $\kappa$ &
$0.1346,0.1351,0.1353$ & $0.13344,0.13417,0.13455$\\
$\kappa_{\mathrm{crit}}$ & $0.13581^{+2}_{-1}$ & $0.13525^{+2}_{-1}$\\
$\kappa_{\mathrm{n}}$ & $0.13578^{+2}_{-1}$ & $0.13520^{+1}_{-2}$\\
$\kappa_{\mathrm{s}}$ & $0.13495^{+2}_{-2}$ & $0.13401^{+2}_{-2}$\\
\hline\hline \end{tabular} } 
$\kappa_{\rm crit}$, the value of the
hopping parameter which corresponds to zero quark mass, $\kappa_s$ and
$\kappa_n$ are taken from~\cite{QLHS} where the lattice spacing has
been fixed from the Sommer scale $r_0$~\cite{sommer_r0,wittig_r0}.
Statistical errors are estimated using the bootstrap~\cite{efron} with
$1000$ re-samplings. Unless otherwise specified, all errors listed in
tables are statistical.

Non-perturbative determinations of the improvement coefficients
are available from two groups.  The ALPHA collaboration have
determined the value of
$c_{\mathrm{SW}}$~\cite{alpha_np,alpha_np2} using chiral symmetry
and Ward identities in the Schr\"{o}dinger Functional (SF)
formalism.  They have determined $c_A$~\cite{alpha_np2} and $Z_A$,
$Z_V$ and $b_V$~\cite{alpha_np3_Z} in the same scheme and have a
preliminary determination of $c_V$~\cite{sommer_cv,alpha_np5_cv}.
Bhattacharya \emph{et al}.~\cite{bhatta_plb,bhatta_99,Bhatta_2000}
have determined all the improvement coefficients needed to improve
and renormalise quark bilinears, also using Ward identities, but
on a periodic lattice with standard sources. They also use the
ALPHA value of $c_{\mathrm{SW}}$ to improve the action. The values
of $Z_V, Z_A$ and $b_V$ are in good agreement between the two
groups. However, the ALPHA values for $c_V$ and $c_A$ are much
larger than those of Bhattacharya \emph{et al}. In a more detailed
discussion of the improvement coefficients~\cite{np_imp_fb} we
noted that the matrix elements determining leptonic decays are
particularly sensitive to the value of the mixing $c$
coefficients. As shown in equation (\ref{eqn:imp_current}) it is
the derivative of a current which is the improvement term. For the
decay constants calculated at zero three-momentum the derivative
is the mass of the state which is $\order{1}$. For the matrix
elements determined here, the momentum component is a momentum
difference and is $\order{1/10}$. Thus these matrix elements are
relatively insensitive to the values of the mixing coefficients.
We use the value of Bhattacharya \emph{et al}. The value of $b_m$
is determined in lattice perturbation theory~\cite{alpha_np4_bA}
with the boosted coupling $g^2=g_0^2/u^4_0$~\cite{lepage_mack}.
The mean link, $u_0$, is taken from the plaquette expectation
value, $u_0^4=\langle\mathrm{Re\,Tr}\, U_{\mathrm P}\rangle/3$.
The values of the improvement coefficients are shown in
table~\ref{tab:imp_coeff}.

\TABLE{
\caption{Improvement coefficients used in this work.\smallskip}
\label{tab:imp_coeff}
\begin{tabular}{cr@{.}lr@{.}l}
\hline\hline
&\multicolumn{2}{c}{$\beta=6.2$} & \multicolumn{2}{c}{$\beta=6.0$}\\
$c_{SW}$ & $1$&$614$ & $1$&$769$ \\
$Z_A$ & $0$&$818(2)(5)$ & $0$&$807(2)(8)$ \\
$b_A$ & $1$&$32(3)(4)$ & $1$&$28(3)(4)$ \\
$c_A$ & $-0$&$032(3)(6)$ & $-0$&$037(4)(8)$ \\\hline
$Z_V$ & $0$&$7874(4)$ & $0$&$770(1)$ \\
$b_V$ &  $1$&$42(1)(1)$ & $1$&$52(1)$ \\
$c_V$ & $-0$&$09(2)(1)$ &  $-0$&$107(17)(4)$ \\\hline
$b_m$ & $-0$&$6517$ & $-0$&$6621$ \\
\hline\hline
\end{tabular}
}

\subsection{Correlation functions}

Details of the fitting procedure, and extraction of the spectrum
and amplitudes between the vacuum and meson states at zero
three-momentum: $\langle M_P | \psb \gamma_5 \psi | 0 \rangle $
for example can be found in \cite{QLHS,np_imp_fb}. We used the
fuzzing smearing function~\cite{fuzzing} for the light quark
propagators and a hydrogenic gauge invariant
function~\cite{boyling_p} for the heavy quarks.

The light-light $2$-point correlation functions at non-zero
momentum were somewhat noisy. To determine these as accurately as
possible we used a constrained dispersion relation fit. That is,
we constrained the energy of the non-zero momentum states to be
 \be \label{eqn:dispersion}
  E^2(|\vec{p}|^2) = M^2 + \vec{p}\cdot\vec{p}
 \ee
Furthermore, we constrained the amplitudes by fitting the correlator
which was fuzzed at source and local at sink (FL) to the following form
 \be
  C(\vec{p},t)=\frac{Z^F(|\vec{p}|^2) Z^L }{2E}
    \left( e^{-Et} + e^{-E(T-t)} \right)
 \ee
where $T$ is the time extent of the lattice and $E$ is constrained
from equation~(\ref{eqn:dispersion}). The local $Z^L$, by Lorentz
invariance, is not a function of momentum and so both $Z^L$ and
$M$ are determined at zero momentum.

We determined the amplitudes and energies of the heavy-light
mesons at all momenta using a free single exponential fit. We then
compared to the dispersion relation as a check of lattice
artefacts.

We have extracted the matrix element for the semileptonic decays
from three-point and two-point correlation functions. The
three-point function for the transition from state $A$ to state $B$
is given by, \ba
   \label{eqn:C3ptop}
  C_{3\rm pt}^{\mu}(\vec{p},t_x,\vec{k},t_y) &=&
  \sum_{\vec{x},\vec{y}}e^{-i(\vec{k}\cdot\vec{x} +
    (\vec{k}-\vec{p})\cdot\vec{y})}
   \langle0|\,\cl{T} \{ \Omega_B(x) J^{\mu}(y)
    \Omega_A^{\dag}(0) \}\, |0\rangle
 \ea
where $\Omega_A^\dag$ is the operator which creates state $A$,
$\Omega_B$ is the operator which destroys state $B$ and $J^{\mu}$
is the weak current. We have computed the correlation function
using the ``standard source'' method~\cite{ext_prop,ext_prop2}. We
chose the extension timeslice to be, $t_x=28$ rather than in the
centre of the lattice at $t=24$. This enabled us to look for
different systematic effects by comparing both sides of the
lattice, for example to look at different time orderings of
operators and to check the quality of plateaux. However, we lost
the ability to average results from the front and back sides of
the lattice and so our statistical errors are larger.

When the operators are well separated, that is when $t_y$ and
$t_x-t_y$ are large, then
 \be
  C_{3\rm pt}^{\mu}(\vec{p},t_x,\vec{k},t_y) =
    \frac{e^{-E_Bt_y}}{2E_B}
   Z_A\frac{e^{-E_\rho(t_x-t_y)}}{2E_\rho}Z_B^\star
    \langle \rho(\vec{k}) | J^{\mu}(0) | \bar{B}(\vec{p})\rangle
 \ee
where $Z_i=\langle 0 | P_i | P_i(\vec{p_i}) \rangle$. Both the
$Z$'s (for smeared operators) and the energies are functions of
three-momentum squared, $|\vec{p}_i|^2$, and can be extracted from
the appropriate two-point functions, as described above. The
remaining matrix element is the one in equation~(\ref{eqn:FF}).

The simulation used heavy meson spatial momentum of magnitude $0$
or $1$, in lattice units of $2\pi/aL$.  The light meson was given
spatial momentum of $0$ or $1$ in lattice units.  Six momentum
channels with different values of $q^2$ were considered, shown in
table~\ref{tab:mom_chan}.

 \TABLE{ \caption{Values of three-momentum in lattice units of
 $\frac{2\pi}{aL}$ and the number of distinct matrix elements for
 each momentum channel. \lq\lq+ perms" implies permutations of
 $\vec{k}$ that have the same $q^2$, }
 \label{tab:mom_chan}
 \begin{tabular}{crrlcc}
 \hline\hline
channel& $\vec{p}$ & $\vec{k}$& &$ \langle \rho|A_\mu|B\rangle$ & $\langle \rho|V_\mu|B\rangle$ \\\hline
0 & $(0,0,0)$ & $ (0,0,0)$ && 1 & 0\\
1 & $(0,0,0)$ & $ (1,0,0)$ &+perms &5 & 1\\
2 & $(1,0,0)$ & $ (1,0,0)$ &&3 & 2\\
3 & $(1,0,0)$ & $ (0,0,0)$ &+perms& 4 & 2\\
4 & $(1,0,0)$ & $ (0,1,0)$ &+perms&10 & 3\\
5 & $(1,0,0)$ & $ (-1,0,0)$&&5 & 1 \\
\hline
\end{tabular}
}

For each channel, different spatial and temporal components
combined with the relevant three-momentum allow
equations~(\ref{eqn:FF}) to be decomposed into distinct matrix
elements. The three-point correlation functions were fitted
simultaneously to the distinct matrix elements with the form
factors as free parameters. We averaged over momenta and Lorentz
channels that have the same matrix element. The number of distinct
matrix elements for each channel and current is shown in
Table~\ref{tab:mom_chan}.  For channel $0$, the coefficients
vanish for all the form factors except $A_1$.

Figure~\ref{fig:ratios} shows examples of the averaged three-point
correlation functions with the time dependence divided out. The
data on the fore-side of the lattice has larger statistical
errors. This is simply due the noise-to-signal ratio increasing at
larger time separations. We observed clear plateaux, but note that
the plateaux on the back-side of the lattice are better for
extracting the signal.  We chose to fit to the three-point
function with the time-dependence cancelled using fitted two point
functions rather than use the ratio of three-point over two-point
correlation functions as the current operator has less overlap
with excited states. This is supported by the data, as we see
longer plateaux when using the fitted two-point parameters.

\FIGURE{
\epsfig{file=plats1_all_346034602000.eps,width=0.45\textwidth,height=0.45\textwidth}
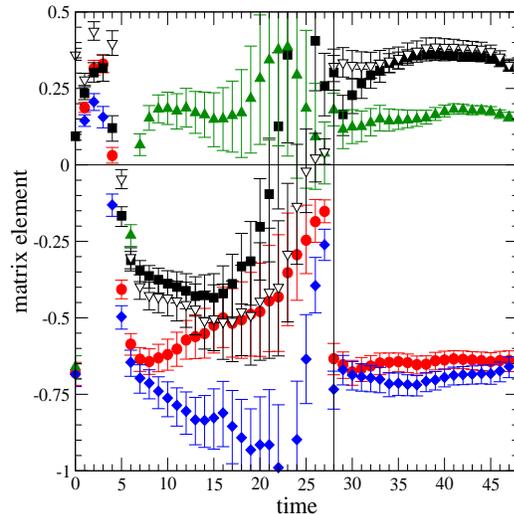
\caption{$C_{3\rm pt}$ with measured two-point factors divided
off, for
         $\beta=6.2$, light quarks, $\kappa_A=\kappa_P=0.1346$ and the
         heavy quark,  $\kappa=0.1200$, kinematic  channel 1.
     Different symbols show the distinct matrix elements for
     $\langle V  | A | P \rangle$.}
\label{fig:ratios}
}

It is clear that even on the shorter back-side of the lattice the
separation of the operators is sufficient to see only the ground
state. For the spatial current (open symbols) there is a
significant disagreement between each side of the lattice. This
may be due to contamination from different time orderings of the
operators wrapping around the lattice. In this case, one would
expect the shorter back-side of the lattice to be less affected
than the longer fore-side.

In general, we fitted to the central five timeslices on the back
side of the lattice. For several channels we have shifted this
window to accommodate a better plateau. Some channels had much
clearer plateaux on the fore-side of the lattice. In all cases the
criteria for choosing the fit range was a plateau such that the
form factors were stable against a change of one timeslice in the
fit range and that had a reasonable reduced $\chi^2$ and
$Q$-value. We used a block diagonal covariance matrix to prevent
different systematics between different currents being
misinterpreted as correlation information.


By measuring the quark mass dependence of the form factors, we can
interpolate and/or extrapolate to the quark masses of the physical
decays we are interested in.

\subsection{Light quark dependence}

In quenched lattice QCD calculations, light quarks are typically
simulated around the strange quark mass and then extrapolated to
light or zero quark mass. The reasons for this are mainly
algorithmic and computational. In the first instance, the
computation of light quark propagators suffers from critical
slowing down. The number of iterations it takes to invert the
fermion matrix varies inversely with the quark mass, making light
quarks prohibitively expensive in computer time. Second, as the
quarks become lighter the hadron states become larger, requiring a
larger box size, again increasing the amount of computation
required. Third, quenched QCD has the wrong chiral behaviour, but
in practice it is the computational requirements which restrict
the range of quark mass.

The form factor $F$ has both explicit, and implicit mass dependence,
 \be
  F=F\left(q^2(m_A,m_P),m_A,m_P\right)
 \ee
where $m_A$ is the active quark mass and $m_P$ is the passive quark mass.
For small changes in $m_A$, $m_P$, and $q^2$ the variation in the form factor
can be approximated by a first-order Taylor expansion
 \be
 \label{eqn:taylor_qmass}
 F(q^2,m_A,m_P) = \alpha +
 m_A\frac{\partial F}{\partial m_A} +
 m_P\frac{\partial F}{\partial m_P} +
 q^2\frac{\partial F}{\partial q^2}
 \ee
For a particular momentum channel the exact dependence of $q^2$ on meson
mass is known.  The first-order Taylor expansion of $q^2$ gives
 \be
 q^2=\beta+\gamma M_H+\delta M_L
 \ee
where $M_H$ is the mass of the heavy pseudoscalar meson and $M_L$
is the mass of the light state.  To a good approximation $M_H$
varies linearly with $m_P$.  The light vector meson mass depends
linearly on the averaged light quark mass
$m_\mathrm{eff}=(m_A+m_P)/2$.

To control this extrapolation, we separated the chiral and momentum
behaviour of the form factors by first interpolating in $q^2$, at
fixed quark mass, to a common set of $q^2$ values over the range
of quark masses. Then we extrapolated in quark mass at fixed
$q^2$, as suggested by~\cite{Lellouch_disp}. To interpolate the
form factors in $q^2$, we used pole
models~\cite{bauer_1,bauer_2,bauer_3},
 \be \label{eqn:pole_model}
  F(q^2) = \frac{F(0)}{1-q^2/m_{F}^2}
 \ee
This is illustrated in figure~\ref{fig:FF_pole_fit}. As we have
interpolated in $q^2$ any model dependence will be small. We have
checked for residual model dependence by varying the model. The
figure shows the form factors and the pole mass fits. We note that
before any heavy quark extrapolation we can cover the full range
of physical $q^2$ values.

 \FIGURE{
\epsfig{file=B2RhoH2000L3460.eps,width=0.6\textwidth,height=0.6\textwidth}
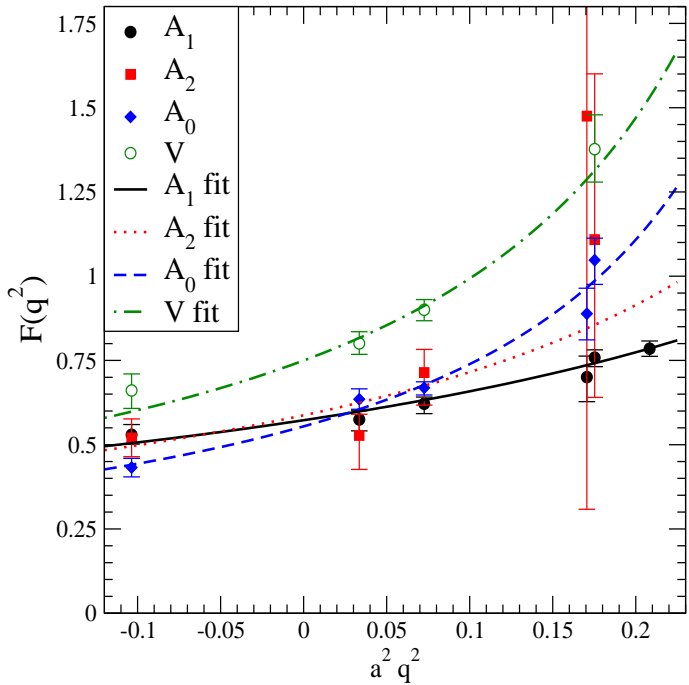
\caption{The form factors as a function of $q^2$, at $\beta=6.2$
for
    $\kappa_H=0.1200$, $\kappa_A=\kappa_P=0.1346$.}
\label{fig:FF_pole_fit}
}

Once the $q^2$ dependence of the form factors had been determined,
we chose a set of common $q^2$ values to which we interpolated at
each value of the quark mass. We then extrapolated the form
factors at fixed $q^2$ in two dimensions according to
 \be
  F(m_{\rm eff},m_P)\Big|_{q^2=\mathrm{const}} =
   \alpha + \beta m_P + \gamma m_\mathrm{eff}
 \ee

\subsection{Heavy quark extrapolation}

Near zero-recoil ($q^2_\mathrm{max}$) the form factors scale in
the following way~\cite{H_scaling},
 \be
 F\sim M^{N/2}
 \ee
where $N=1$ for $F\in\{V,A_0,A_2\}$, $N=-1$ for $F=A_1$. We
introduce a new kinematic variable
 \be
  \vk=\frac{M_H^2+M_L^2-q^2}{2M_H}
 \ee
where $v$ is the four-velocity of the heavy meson. We then have the following
form for the extrapolation at fixed $\vk$.
 \be
 \label{eqn:HQS_extrap}
  \Phi_i\equiv C\left(M_H,M_B\right)F(\vk)M^{-N/2}=\zeta\left(1 +
      \frac{\eta}{M_H} + \frac{\theta}{M_H^2} + \cdots \right)
 \ee
The coefficient $C$ is the logarithmic matching factor~\cite{neubert_coeff},
\be
  C\left(M_H,M_B\right)=\left(\frac{\alpha_s(M_B)}{\alpha_s(M_H)}\right)^{
    2/\beta_0}
 \ee
where $\beta_0=11$ in quenched QCD and $\alpha_s$ is the one-loop
running coupling with $\Lambda^{(4)}_{\overline{MS}}=295\mev$.

With the fixed-$q^2$ extrapolation, it is easy to implement the
heavy quark extrapolation. We used a different set of $q^2$ at
each value of the heavy quark mass so that the heavy quark
extrapolation could be done at fixed $\vk$.

\FIGURE{
\epsfig{file=B2rhoheavyExtrap.eps,width=0.6\textwidth,height=0.6\textwidth}
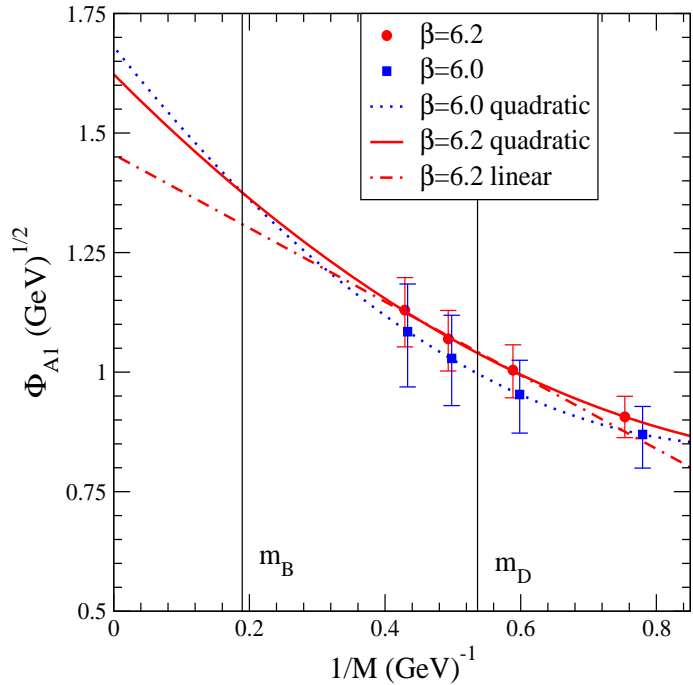
\caption{The heavy quark extrapolation for the form factor $A_1$
at the highest
         value of $q^2$.}
\label{fig:heavyExtrap}
}

Figure~\ref{fig:heavyExtrap} shows the extrapolation of the form
factor $A_1$. The solid line shows the quadratic fit to all four
data points at $\beta=6.2$, and the dot-dashed line shows the
linear fit to the heaviest three. At the $B$ meson mass the two
lines are still close together, suggesting that the extrapolation
is reasonable. Comparing the circles ($\beta=6.2$) with the
squares ($\beta=6.0$) there is little difference, suggesting that
$\order{(am)^2}$ lattice artefacts are small. In~\cite{np_imp_fb}
we fitted simultaneously to the data for the decay constants at
both lattice spacings, allowing for lattice artefacts, and found
the effect to be small. The method of extrapolation is the same as
used here. However, we do not attempt a simultaneous fit since the
statistical precision of the data is not sufficiently good even
for $A_1$.  

\section{Results}

In principle one could compute the form factors for any value of
$q^2$ from lattice QCD. However, states with high spatial momentum
are very noisy and thus difficult to measure on the lattice. Thus
we are restricted to the high $q^2$ end of the range. In addition,
the procedures we have introduced to control the extrapolations,
separating the $q^2$ from the quark mass dependence, have further
restricted the range of $q^2$ away from $q^2_\mathrm{max}$, in the
range
 \be \label{eqn:qSqrRange}
  12.7\gev^2 \le q^2 \le 18.2\gev^2
 \ee
Moreover, the relatively small number of momentum channels for
which the form factors are extracted, six for $A_1$, five for
$A_0$ and $A_2$, and four for $V$, coupled with the interpolation
at fixed $q^2$ imply by naive counting of degrees of freedom that
we have only four independent data for $A_1$ and worse, two
independent data for $V$. Fitting the functional form of the $q^2$
dependence of the form factors is thus rather hard. However, we
are free to evaluate the form factors, and thus the differential
decay rate, at any value of $q^2$ we choose without introducing
any extra model dependence as long as it is in the range of
allowed $q^2$. In particular we can determine a partially
integrated decay rate over this range.

Figure~\ref{fig:B2RhoFF} shows the four form factors on both
lattices. In this case we have chosen nine values of $q^2$. The
form factor $A_1$ which dominates at $q^2_\mathrm{max}$ is well
determined and is in good agreement for both lattice spacings. The
subleading form factors have a much noisier signal, especially for
the coarser lattice.  This made the extrapolations very difficult
to control. For the coarse lattice only we introduced two
additional model dependent constraints on the data during the fixed
quark mass $q^2$ interpolation. First, according to pole models
the pole mass for $A_1$ and $A_2$ should be the same, and we
enforce this constraint. As $A_1$ is much better determined than
$A_2$, $A_1$ remains unchanged. Second, we find a dipole rather
than a pole fits the data better for $V$. Both these constraints
affect the value of the form factors very little for the fine
lattice and are not required to control the extrapolations. As the
finer lattice forms our result and the coarse lattice is used to
estimate lattice artifacts, the resulting model dependence enters
only in our estimate of systematic error.

\FIGURE{
\epsfig{file=B2RhoFF.eps,width=0.6\textwidth,height=0.6\textwidth}
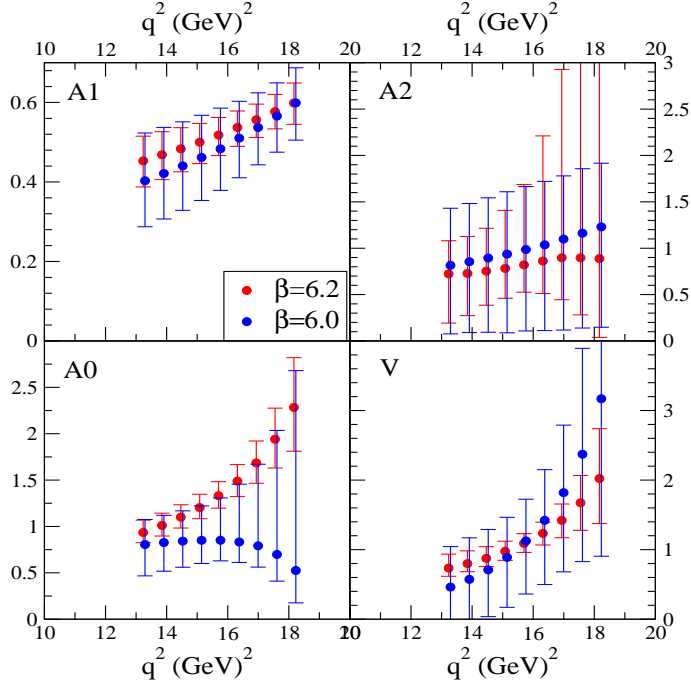
\caption{The form factors on both lattices. The vertical scale is
different for each form factor.} \label{fig:B2RhoFF} }

\TABLE{
\caption{The form factors on both lattices}
\label{tab:B2RhoFF}
\begin{tabular}{ccc}
$q^2$ (GeV)${}^2$ & $\beta=6.2$ & $\beta=6.0$ \\ \hline
& \multicolumn{2}{c}{$A_1$} \\
$18.2$ & $0.60^{+5}_{-5}$ & $0.60^{+\ 9}_{-\ 9}$\\
$12.7$ & $0.45^{+6}_{-7}$ & $0.40^{+12}_{-12}$\\\hline
& \multicolumn{2}{c}{$A_2$} \\
$18.2$ & $0.88^{+500}_{-\ 85}$ & $1.23^{+\ 69}_{-108}$ \\
$12.7$ & $0.72^{+\ 36}_{-\ 53}$ & $0.82^{+\ 62}_{-\ 74}$ \\ \hline
& \multicolumn{2}{c}{$A_0$} \\
$18.2$ & $2.28^{+53}_{-47}$ & $0.52^{+216}_{-\ 35}$ \\
$12.7$ & $0.93^{+13}_{-11}$ & $0.80^{+\ 27}_{-\ 34}$ \\ \hline
& \multicolumn{2}{c}{$V$} \\
$18.2$ & $2.02^{+72}_{-65}$ & $3.17^{+247}_{-226}$ \\
$12.7$ & $0.74^{+20}_{-12}$ & $0.46^{+\ 58}_{-\ 58}$ \\ \hline
& \multicolumn{2}{c}{$\frac{\partial \Gamma}{\partial q^2}\ \times 10^{12} \ s^{-1}$} \\
$18.2$ & $1.4^{+\ 5}_{-\ 5}$ & $1.6^{+16}_{-\ 5}$\\
$12.7$ & $1.1^{+10}_{-\ 4}$& $0.6^{+\ 9}_{-\ 1}$\\\hline
\end{tabular}
}

In Table~\ref{tab:B2RhoFF} we list the values of the form factors
on both lattices at the highest and lowest values of $q^2$ only.
Using equation~(\ref{eqn:ddr_V}) we can similarly determine the
differential decay rate over the same range of momentum. Shown in
figure~\ref{fig:B2RhoDDR} is the differential decay rate on both
lattices. To reiterate, we can evaluate the differential decay
rate for \emph{any} $q^2$ in the allowed range but only two data
points will be independent.  For this reason, our main result is
for the partially integrated decay rate over the allowed range
given in equation~(\ref{eqn:qSqrRange})
 \be
 \label{eqn:PIDDR}
  \Gamma_\mathrm{PI} =
   \int_{12.7\gev^2}^{18.2\gev^2}
    \!\! dq^2 \frac{\partial\Gamma}{\partial q^2} =
   \GammaPI\,|\vub|^2,
 \ee
where the first error is statistical and the second is systematic.
Estimates of systematic error are discussed in the next section.

\FIGURE{
 \epsfig{file=B2RhoDDR.eps,width=0.6\textwidth,height=0.6\textwidth}
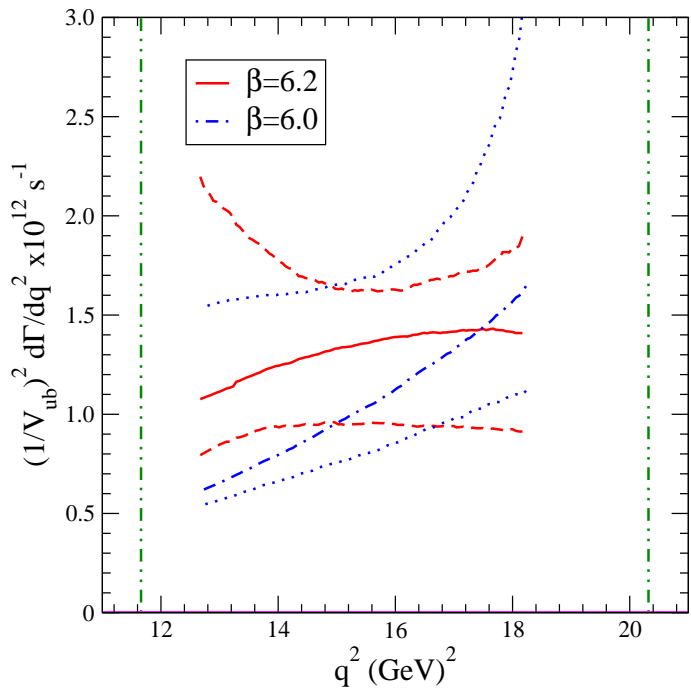
 \caption{The differential decay rate on both lattices. The
 vertical line to the left is the charm end point, and to the right
 is $q^2_\mathrm{max}$.}
 \label{fig:B2RhoDDR}}

The CLEO collaboration~\cite{CLEO_B2PiRho} have measured the rates
for the decays $B\to \pi l\nu$ and $B\to \rho l\nu$. They
determine the partial decay rate for three ranges of $q^2$: $0 <
q^2 < 8\gev^2$, $8< q^2 < 16\gev^2$, and $ q^2 \ge 16\gev^2$. They
compare the measured partially integrated decay rate with Light
Cone Sum Rules (LCSR) results for $q^2<16\gev^2$, and to lattice
QCD data for $q^2>16\gev^2$. We cannot bin our data in the same
way as we have not determined the form factors at
$q^2_\mathrm{max}$, so we cannot directly compare. The BaBar
collaboration~\cite{BABAR_B2Rho} have measured the total decay
rate, but only quote for the lifetime and $\vub$, so again we
cannot compare our results to the experimental data at the same
values of $q^2$.

\subsection{Systematic uncertainties}

This calculation is done in the quenched approximation. There is
therefore an uncontrolled systematic error. For most dimensionful
quantities such as the hadron spectrum this is of the order of
$10\%$. The quenching error is manifest in the choice of quantity
used to set the scale: in the quenched approximation different
quantities such as the nucleon mass, or the Sommer scale, give
different answers for the lattice spacing. The form factors are
dimensionless and so are not directly affected by the
scale-setting. However there is an indirect dependence. We vary
the choice of quantity used to set the scale. This in itself is
not an estimate of the quenching systematic on the form factors,
only of their implicit scale dependence.  We use the Sommer scale,
$r_0$ to set the scale. At each lattice spacing the value of
$a/r_0$ is unambiguously defined although the experimental value
is not known. Sommer originally advocated $r_0=0.5\fm$ but
determinations of the lattice spacing using the kaon decay
constant, the nucleon mass or the rho mass correspond to $r_0$
values in the range $0.5$--$0.55\fm$. We take $r_0=0.5\fm$ as our
central value and use $r_0=0.55\fm$ to estimate the systematic
error from quenching scale ambiguity.

We simulate heavy quarks with several quark masses around the
charm quark mass and then use continuum heavy quark symmetry to
extrapolate the form factors to the bottom quark mass. We estimate
the systematic uncertainty coming from the quadratic extrapolation
in equation~(\ref{eqn:HQS_extrap}) by performing a linear
extrapolation to the heaviest three quark masses. This is shown as
the dot-dashed line in figure~\ref{fig:heavyExtrap}. For the form
factor $A_1$ on the finer lattice spacing at high $q^2$ this
effect is smaller than the statistical uncertainty. For the other
form factors the effect is larger, so we include this in the
systematic uncertainty of the partially integrated decay rate.

In~\cite{np_imp_fb} we made several estimates of systematic
uncertainty arising from the heavy-quark formalism and
renormalisation method.  We argue here that the same effects will
be no larger for the form factors than for the heavy-light decay
constants. We simultaneously fitted the heavy-light decay constant
data at both lattice spacings and allowed for mass dependent
lattice artefacts. This produced a $5\%$ effect in the value of
decay constant. We do not attempt this procedure here as our
statistical errors are too large, but have no reason to believe it
would be a greater effect for the form factors. As suggested by
Bernard~\cite{Bernard_2000} the normalisation of the currents can
be altered so that while formally the same at $\order{a}$ it looks
much more like the mass dependent normalisation of the Fermilab
formalism~\cite{KLM_norm}.
 \be
  \psi \to \psi^{\prime} = \sqrt{1+\mu am}\,\psi
 \ee
where $\mu$ depends on the improvement coefficients $Z_J$, $c_J$
and $b_J$ and the improved currents defined in
equation~(\ref{eqn:imp_current}). Combining this alternative
normalisation with the kinetic mass $M_2$ defined as
 \be
 \frac{1}{M_2}=\frac{\partial^2 E}{\partial p_k^2}\Bigg|_{\vec{p}=0}
 \ee
we found that this produced an effect on the pseudoscalar decay
constant smaller than the difference between the linear and quadratic
extrapolations at the coarser lattice spacing. As we described in
section~\ref{sec:imp-action-ops}, the effect of the improvement
for the form factors is much smaller than for the decay constants,
so the overall effect of changing the normalisation and hadron
mass cannot be greater than for the decay constants. Finally we
varied the values of the improvement coefficients by the errors
quoted by Bhattacharya \emph{et al}.~\cite{bhatta_plb} and found
the overall effect to be $1\%$ in the axial and $3\%$ in the
vector currents. Again, as the effect of improvement is smaller in
the form factors the overall effect cannot be greater than this.

The main systematic uncertainties are tabulated in
table~\ref{tab:sysError}. To obtain the systematic error on the partially
integrated decay rate we have combined the systematic errors in
quadrature.

\TABLE{
\caption{Systematic uncertainties in the decay rate}
\label{tab:sysError}
\begin{tabular}{lccc}
& $\Gamma_{\rm PI}$ & $\frac{\partial \Gamma}{\partial q^2}|_{\rm low}$&
  $\frac{\partial \Gamma}{\partial q^2}|_{\rm high}$\\\hline
lattice spacing &$-20\%$& $-45\%$ & $+14\%$ \\
quenched scale ambiguity &$-20\%$&$-27\%$&$-14\%$\\
heavy quark extrapolation &$-16\%$&$-9\%$&$-14\%$\\
pole model dependence &$-4\%$&$-5\%$&$+0\%$\\
\end{tabular}
}

\newpage
\acknowledgments
We thank PPARC for support under grants
PPA/G/O/2002/00468, 
PPA/G/O/2002/00465 
and PPA/G/S/2002/00467. 
We acknowledge EPSRC for support under grant
GR/K41663 


\begin{thebibliography}{10}

\bibitem{wolfenstein}
L. Wolfenstein, Phys.~Rev.~Lett. {\bf 51},  1945  (1983).

\bibitem{Belle_s2b_2003}
{Belle Collaboration, K.~Abe } {\it et~al.}, \texttt{hep-ex/0308036}.

\bibitem{BaBar_s2b_2002}
{BABAR Collaboration, B.~Aubert} {\it et~al.}, Phys.~Rev.~Lett. {\bf 89},
  201802  (2002).

\bibitem{korner_schuler_dr}
{J.G.~K\"{o}rner and G.A.~Schuler}, Phys.~Lett.~B {\bf 231},  306  (1989).

\bibitem{gilman_singleton}
{F.J.~Gilman and R.L.~Singleton~Jr.}, Phys.~Rev.~D {\bf 41},  142  (1990).

\bibitem{sw_paper}
B. Sheikholeslami and R. Wohlert, Nucl.~Phys.~B {\bf 259},  572  (1985).

\bibitem{KLM_norm}
{A.X.~El-Khadra, A.S.~Kronfeld, and P.B.~Mackenzie}, Phys.~Rev.~D {\bf 55},
  3933  (1997).

\bibitem{B2pi_plb}
{UKQCD Collaboration, K.C.~Bowler} {\it et~al.}, Phys.~Lett.~B {\bf 486},  111
  (2000).

\bibitem{ukqcd_gill_brho}
{UKQCD Collaboration, J.~Gill}, Nucl.~Phys.~B~(Proc.~Suppl.) {\bf 106},  391
  (2002).

\bibitem{creutz_or}
M. Creutz, Phys.~Rev.~D {\bf 36},  2394  (1987).

\bibitem{brown_woch_or}
F. Brown and T. Woch, Phys.~Rev.~Lett. {\bf 58},  163  (1987).

\bibitem{cabibbo_marinari}
N. Cabibbo and E. Marinari, Phys.~Lett.~B {\bf 119},  387  (1982).

\bibitem{QLHS}
{UKQCD Collaboration, K.C.~Bowler} {\it et~al.}, Phys.~Rev.~D {\bf 62},  054506
   (2000).

\bibitem{sommer_r0}
R. Sommer, Nucl.~Phys. {\bf B411},  839  (1994).

\bibitem{wittig_r0}
{M.~Guagnelli, R.~Sommer and H.~Wittig}, Nucl.~Phys.~B {\bf 535},  389  (1998).

\bibitem{efron}
B. Efron, SIAM Review {\bf 21},  460  (1979).

\bibitem{alpha_np}
{M.~L\"{u}scher, S.~Sint, R.~Sommer and P.~Weisz}, Nucl.~Phys.~B {\bf 478},
  365  (1996).

\bibitem{alpha_np2}
{M.~L\"{u}scher, S.~Sint, R.~Sommer, P.~Weisz, H.~Wittig and U.~Wolff},
  Nucl.~Phys.~B~(Proc.~Suppl.) {\bf 53},  905  (1997).

\bibitem{alpha_np3_Z}
{M.~L\"{u}scher, S.~Sint, R.~Sommer, and H.~Wittig}, Nucl.~Phys.~B {\bf 491},
  344  (1997).

\bibitem{sommer_cv}
R. Sommer, Nucl.~Phys.~Proc.~Suppl {\bf 60A},  279  (1998).

\bibitem{alpha_np5_cv}
M. Guagnelli and R. Sommer, Nucl.~Phys.~B~(Proc.~Suppl.) {\bf 63A-C},  886
  (1998).

\bibitem{bhatta_plb}
T. Bhattacharya {\it et~al.}, Phys.~Lett.~B {\bf 461},  79  (1999).

\bibitem{bhatta_99}
{T.~Bhattacharya, R.~Gupta, W.~Lee, S.~Sharpe}, Nucl.~Phys.~B~(Proc.~Suppl.)
  {\bf 83-84},  851  (2000).

\bibitem{Bhatta_2000}
{T.~Bhattacharya, R.~Gupta, W.~Lee, S.~Sharpe}, Phys.~Rev.~D {\bf 62},  074505
  (2001).

\bibitem{np_imp_fb}
{UKQCD Collaboration, K.C.~Bowler} {\it et~al.}, Nucl.~Phys.~B {\bf 619},  507
  (2001).

\bibitem{alpha_np4_bA}
{S.~Sint and P.~Weisz}, Nucl.~Phys.~B {\bf 502},  251  (1997).

\bibitem{lepage_mack}
G. Lepage and P. Mackenzie, Phys.~Rev.~D {\bf 48},  2250  (1993).

\bibitem{fuzzing}
{P.~Lacock, A.~McKerrell, C.~Michael, I.M.~Stopher and P.W.~Stephenson},
  Phys.~Rev.~D {\bf 51},  6403  (1995).

\bibitem{boyling_p}
P. Boyle, J.~Comput.~Phys. {\bf 179},  349  (2002).

\bibitem{ext_prop}
{C.~Bernard and A.~Soni}, Nucl.~Phys.~B~(Proc.~Suppl.) {\bf 9},  155  (1989).

\bibitem{ext_prop2}
{C.~Bernard},  in {\em Gauge Theory on a Lattice: 1984}, {\em Argonne National
  Laboratory Workshop}, edited by {C.~Zachos} {\it et~al.} (National Technical
  Infomation Service, Springfield, VA, 1984).

\bibitem{Lellouch_disp}
{L.~Lellouch}, Nucl.~Phys.~B {\bf 479},  353  (1996).

\bibitem{bauer_1}
{M.~Bauer, B.~Stech and M.~Wirbel}, Z.~Phys.~C {\bf 29},  627  (1985).

\bibitem{bauer_2}
{M.~Bauer, B.~Stech and M.~Wirbel}, Z.~Phys.~C {\bf 34},  103  (1987).

\bibitem{bauer_3}
{M.~Bauer and M.~Wirbel}, Z.~Phys.~C {\bf 42},  671  (1989).

\bibitem{H_scaling}
N. Isgur and M. Wise, Phys.~Rev.~D {\bf 42},  2388  (1990).

\bibitem{neubert_coeff}
M. Neubert, Phys.~Rev.~D {\bf 49},  1542  (1994).

\bibitem{CLEO_B2PiRho}
{CLEO Collaboration, S.B.~Athar} {\it et~al.}, Phys.~Rev.~D {\bf 68},  072003
  (2003).

\bibitem{BABAR_B2Rho}
{BaBar Collaboration, B.~Aubert} {\it et~al.}, Phys.~Rev.~Lett. {\bf 90},
  181801  (2003).

\bibitem{Bernard_2000}
{C.~Bernard}, Nucl.~Phys.~B~(Proc.~Suppl.) {\bf 94},  159  (2001).

\end{thebibliography}

\end{document}